\documentclass[a4paper,aps,prl,onecolumn,a4paper]{revtex4}
\usepackage{epsfig}
\usepackage{graphicx}
\usepackage{epsf}
\usepackage{bm}
\usepackage[T1]{fontenc}
\usepackage{amsmath,amssymb}
\usepackage[latin1]{inputenc}
\usepackage{times}
\usepackage{ulem}
\usepackage{color}
\usepackage[squaren,Gray]{SIunits}

\usepackage[squaren,Gray]{SIunits}
\usepackage{times}

\usepackage{amsmath}
\usepackage{amsfonts}
\usepackage{amssymb}
\usepackage{graphicx}
\usepackage[latin1]{inputenc}
\usepackage{color}

\begin{document}
\title{Multiple scattering limit in optical microscopy}

\author{Amaury Badon, A. Claude Boccara, Geoffroy Lerosey, Mathias Fink, Alexandre Aubry}
\email{alexandre.aubry@espci.fr}

\affiliation{ESPCI Paris, PSL Research University, CNRS, Institut Langevin, UMR 7587, 1 rue Jussieu, F-75005, Paris, France }




\begin{abstract}
Optical microscopy offers a unique insight of biological structures with a sub-micrometer resolution and a minimum invasiveness. However, the inhomogeneities of the specimen itself can induce multiple scattering of light and optical aberrations which limit the observation to depths close to the surface. To predict quantitatively the penetration depth in microscopy, we theoretically derive the single-to-multiple scattering ratio in reflection. From this key quantity, the multiple scattering limit is deduced for various microscopic imaging techniques such as confocal microscopy, optical coherence tomography and related methods.
\end{abstract}

\maketitle



\section{Introduction}

For decades, optical imaging has been a vital tool in biology as it enables to investigate non-invasively living specimens \cite{milestone}. With a diffraction limited resolution of the order of 1 micron, a conventional microscope can reveal sub-cellular structures in biological tissues. However, the imaging depth that can be achieved remains shallow due to the scattering of light that occurs during its propagation inside the specimen. Indeed, the principle of a conventional microscope relies on a weak interaction between the incident light and the specimen which corresponds to the Born approximation. In a reflection configuration, this approximation supposes that the reflected light has been scattered only once by the specimen and experienced a ballistic propagation from the source to the scatterers and once again from the scatterers to the detector. Yet, an high concentration of heterogeneities in the specimen can induce multiple scattering (MS) events that make this assumption no longer valid. Thus, the reflected light can be splitted into two terms: a single scattering (SS) contribution - the component of interest - and a detrimental MS contribution. Typically, beyond one scattering mean free path $\ell_s$, the mean distance between two scattering events, the MS contribution becomes predominant and a conventional microscope fails to yield an accurate image of the medium. The ability to image an inhomogeneous medium is therefore directly given by the ratio of the SS and MS contributions that we will refer to as the single-to-multiple scattering ratio (SMR). It directly provides the image contrast and should not be confused with the signal-to-noise ratio that can also limit the imaging depth in low-sensitivity optical imaging systems.

To cope with the fundamental issue of MS in optical imaging, several strategies have been proposed to enhance the SS contribution and thus to extend the imaging depth \cite{dunsby2003techniques}. The first option is to spatially discriminate SS and MS as performed in confocal microscopy \cite{minsky1955confocal} or two-photon microscopy \cite{2P_denk}. The principle is to select, for each input illumination focused at one location of the sample, only the reflected light that comes from the illumination point.  In practice, either a pinhole conjugated with a point source placed in front of the detector or the high density of power at focus ensure spatial selection. A complete image is then obtained by repeating this scheme for all the positions inside the field-of-view (FOV). One can see this as the measurement of the impulse response of the system between a point source at the input and a point detector at the pinhole location. The second option consists in separating SS from MS photons by means of time gating \cite{Fujimoto2,Alfano}. The most widely used coherence time-gated technique is probably optical coherence tomography (OCT) \cite{fujimoto}, which is analogous to ultrasound imaging. It combines scanning confocal microscopy with coherent heterodyne detection. By taking advantage of a time-resolved impulse response, OCT has drastically extended the imaging-depth limit compared to conventional microscopy. Nevertheless, its ability in imaging soft tissues remains typically restricted to a depth of 1 mm \cite{dunsby2003techniques}.

For larger depths, the MS light is usually considered as randomized and modeled with the diffusion theory which neglects all the interferences effects. However, recent developments in light manipulation techniques demonstrated the possibility to focus light spatially through a strongly scattering medium by properly shaping the incident wavefront \cite{Vellekoop}. Subsequently, a matrix approach to light propagation through such media was proposed \cite{popoff}. Using the knowledge of impulse responses between a spatial light modulator (SLM) and a CCD camera, this transmission matrix allows to take advantage of the MS light to focus at a given position or transmit information across a scattering medium \cite{popoff2010image}. Even more recently, this approach was adapted to the reflection configuration for imaging purposes \cite{Kang}. Compared to OCT, an extension of the imaging depth was reported both experimentally and theoretically \cite{badon2016smart}. As a lot of efforts are currently made to increase the penetration depth in microscopy \cite{Kang,badon2016smart,Park2015,jang2013}, a quantitative prediction of the MS limit in optical imaging is needed. 

In spite of the crucial role of MS in microscopy, only few studies have analytically addressed this issue to our knowledge \cite{Schmitt0,Kempe,Schmitt,Thrane,Karamata,oct,carney2009}. Schmitt \textit{et al.} \cite{Schmitt0} investigated the SS contribution in confocal microscopy but their MS analysis was restricted to a Monte Carlo simulation. Kempe \textit{et al.} \cite{Kempe} analytically derived the SMR for a confocal microscope, but only in a transmission configuration. In OCT, most studies assumed that MS only arises from photons that have been backscattered from the target layer and not from the bulk tissue \cite{Schmitt,Thrane,Karamata,oct}. On the contrary, in this article, we focus on the latter contribution that dominates at large probing depths and gives rise to a predominant diffuse background \cite{Yadlowsky,Yao}. 

In this article, we propose to study the performance of microscopy imaging techniques in reflection. To that aim, we investigate the impulse responses between the source and detector planes, both in the single and multiple scattering regimes. A theoretical expression of the SMR is then derived for different imaging apparatuses: a conventional microscope, a confocal microscope and an OCT system. After a description of the imaging conditions considered throughout this article (section 1), the performances of these imaging techniques are firstly investigated for a point-like target embedded in a turbid medium (sections 3-6). Then, these results will be expanded to an extended target (section 7). Finally, the imaging of the turbid medium itself is addressed by considering the example of biological soft tissues (section 8). 

\section{Imaging conditions: Optical microscope in a reflection configuration}

\begin{figure}[htbp]
\begin{center}
\includegraphics[width=\textwidth]{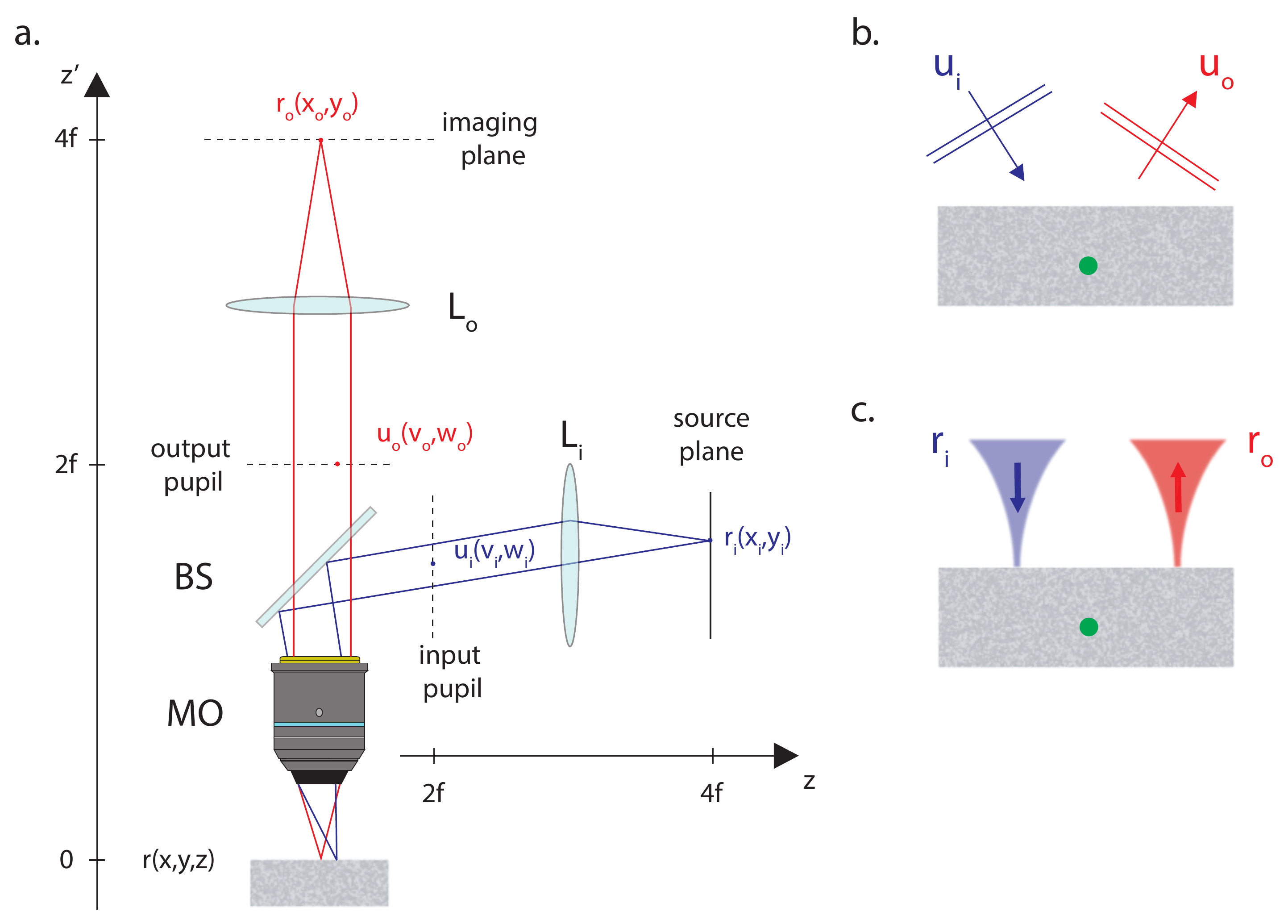}
\caption{ \label{fig1} Imaging configuration. (a) A specimen in the focal plane of a MO is illuminated by a source at position $\mathbf{r_i}$ in the back focal plane of a lens $L_i$. The reflected light is collected through the same MO and measured in the focal plane of a lens $L_o$. (b) Plane wave illumination configuration. (c) Focused illumination configuration.}
\end{center}
\end{figure}

Figure \ref{fig1}(a) describes the system consisting of the microscope investigated throughout this article. 
The specimen that we want to image is placed in the focal plane of a microscope objective (MO) of focal length $f$. The sample is illuminated by a source either at position $\mathbf{u_i}$ in the pupil plane (plane wave illumination) or at position $\mathbf{r_i}$ in the back focal plane of a lens $L_i$ (focused illumination). The reflected light is collected through the same MO and its intensity is measured either at position $\mathbf{u_o}$ in the pupil plane or at position $\mathbf{r_o}$ in the focal plane of a lens $L_o$. For the sake of simplicity but without loss of generality, we will assume that the magnification of the microscope is 1 (4f-system).
As seen in Fig.~\ref{fig1}, input-output responses can connect points located in the pupil planes ($\mathbf{u_i}$ and $\mathbf{u_o}$) or in the source/image conjugate planes ($\mathbf{r_i}$ and $\mathbf{r_o}$, respectively). The pupil plane is associated with a plane wave illumination at emission or a plane wave decomposition of the scattered wave-field at reception [see Fig.~\ref{fig1}(b)]. In the conjugate focal planes, a point source $\mathbf{r_i}$ gives rise to a focused illumination in the focal plane, while a receiver at point $\mathbf{r_o}$ detects light coming from a focal spot in the sample plane. 

\begin{figure}[!ht]
\begin{center}
\includegraphics[width=1\textwidth]{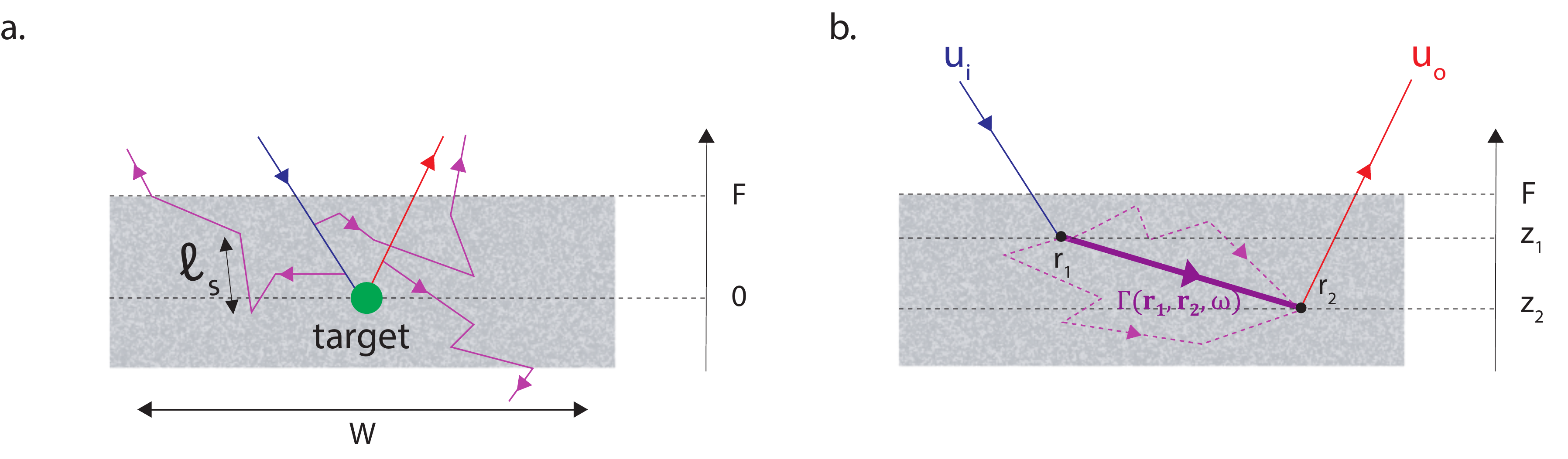}
\caption{ \label{fig2} (a) Imaging of a target embedded at depth $F$ inside a scattering medium of scattering mean-free path $\ell_s$. Both the incident and reflected wave-fields can be scattered by the heterogeneities of the medium before or after reaching the target. (b) Between the first and the last scattering event, the propagation of light is modeled with a propagator $\Gamma$ using the radiative transfer or diffusion theories. }
\end{center}
\end{figure}

As a first step, we will consider, in the next sections, the imaging of a single scatterer, referred to as a target, embedded, at a depth $F$, within a diffusive layer of thickness $L$ [see Fig.~\ref{fig2}(a)]. Wave transport across the scattering layer is described by the usual transport parameters: $\ell_s$, the scattering mean free path, $\ell_t$, the transport mean free path and, $D$, the diffusion constant. Due to the diffusive layer, the impulse response $g$ between the input and output points can be decomposed as the sum of two contributions: A SS contribution $g_S$ associated with the target; A diffusive contribution $g_M$ which corresponds to the set of MS paths induced by the diffusive layer. We now propose to derive the expression of these two contributions in order to express the SMR in various imaging scenarios. This key quantity will give the ability of the imaging system to detect a target embedded within a strongly scattering medium.

\section{\label{sec1}Single-to-multiple scattering ratio in the pupil plane}

In this section, we first want to assess theoretically the SS and MS components of the impulse response for a source/receiver couple in the pupil plane. A monochromatic point source located at point $\mathbf{u_i}=(v_i,w_i)$ in the input pupil plane ($z=2f$) illuminates the sample at frequency $\omega$. A virtual detector records the back-scattered wave-field at point $\mathbf{u_o}=(v_o,w_o)$ in the output pupil plane  [see Fig.~\ref{fig1}(a)]. 

\subsection{Single scattering contribution}

Under the Born approximation, the SS impulse response, $g_S(\mathbf{u_o},\mathbf{u_i},\omega)$, can be expressed as \cite{akkermans}:
\begin{equation}
\label{gS}
g_S(\mathbf{u_o},\mathbf{u_i},\omega)= \int d^3 \mathbf{r}  g_{l}(\mathbf{u_o},\mathbf{r},\omega) v(\mathbf{r}) g_{l}(\mathbf{r},\mathbf{u_i},\omega) 
\end {equation} 
where $v(\mathbf{r})$ is the scattering potential and $g_{l} (\mathbf{r},\mathbf{u_i},\omega)$ is the lens impulse response between the point source $\mathbf{u_i}$ and a point $\mathbf{r}=(x,y,z)$ in the vicinity of the MO focal plane. By reciprocity, $g_{l}(\mathbf{u_i},\mathbf{r},\omega)=g_{l}(\mathbf{r},\mathbf{u_i},\omega)$. 
The ensemble average of the lens impulse response, denoted as $ \langle g_{l} \rangle$, is given by \cite{goodman_fourier}:
\begin{equation}
\label{gL}
\langle g_l(\mathbf{r},\mathbf{u_i},\omega) \rangle=\frac{e^{-\frac{F-z}{2 \mu_i \ell_s }}e^{ik(2f-z)}e^{-i\frac{ k z}{2 f^2} | \mathbf{r_{\|}}|^2}e^{-i \frac{k}{ f}\mathbf{r_{\|}}.\mathbf{u_i}}}{i\lambda f} p(\mathbf{u_i})
\end{equation}
with $\theta_i=\arctan (||\mathbf{u_i}||/f)$, the incident angle, $\mu_i=\cos \theta_i$, its director cosine, $\lambda$, the wavelength, $k=2\pi/\lambda$, the wave number and $\mathbf{r_{\|}}=(x,y)$ the transverse coordinates in the focal plane of the MO. The term $e^{-(F-z)/\ell_s}$ corresponds to the exponential decrease of the ballistic wave across the diffusive layer \cite{akkermans}. $p(\mathbf{u_i})$ is the pupil function \cite{goodman_fourier}. Its modulus accounts for the finite aperture of the MO:
\begin{equation}
\label{aperture}
|p(\mathbf{u_i})|=\left \{
\begin{array}{ll}
1 & \, \mbox{inside the lens aperture}  \\
0 & \,  \mbox{otherwise} 
\end{array} 
\right. 
\end{equation} 
The phase of the pupil function may account for the aberrations induced by the imperfections of the optical system or the scattering medium itself. The target is assumed to be point-like and located at the origin, such that \cite{akkermans}:
\begin{equation}
\label{pointlike}
v(\mathbf{r})=\sqrt{\sigma} \delta(\mathbf{r})
\end{equation}
with $\sigma$, the scattering cross-section of the target and $\delta$, the Dirac distribution. Under this approximation, the ensemble average of Eq.~(\ref{gS}) can be written as:
\begin{equation}
\label{eq2}
\langle g_S(\mathbf{u_o},\mathbf{u_i},\omega) \rangle = -\frac{\sqrt{\sigma}}{(\lambda f)^2} e^{-\frac{F}{ 2 \ell_s } \left (\frac{1}{\mu_i}+\frac{1}{\mu_o}\right) }e^{4ikf}  p(\mathbf{u_o}) p(\mathbf{u_i})~.
\end {equation}
with $\theta_o=\arctan (||\mathbf{u_o}||/f)$, the output angle and $\mu_o=\cos \theta_o$, its director cosine. Because of the point-like feature of the target, the SS intensity $| g_S(\mathbf{u_o},\mathbf{u_i},\omega)|^2$ is constant over the pupil area. 

\subsection{Multiple scattering contribution}

The MS component of the impulse response in the pupil plane, denoted as $g_M(\mathbf{u_o},\mathbf{u_i},\omega)$, is now investigated. Here, the scatterers are assumed to be randomly distributed in the diffusive layer and a probabilistic approach is adopted. In the MS regime, the mean intensity of the impulse response $\left \langle |g_M(\mathbf{u_o},\mathbf{u_i},\omega)|^2 \right \rangle $ is made of an incoherent contribution (diffusive halo) and of a coherent component related to the coherent backscattering effect \cite{albada,wolf1985weak,akkermans,Akkermans2}. For the sake of simplicity, we will neglect the coherent contribution that is angle-dependent. This approximation is justified by the fact that the coherent backscattering effect results in an intensity enhancement by a factor of two only in the vicinity of the backscattering direction. Hence, it will not modify drastically the order of magnitude of the intensity scattered by the diffusive layer. Upon this approximation, the mean intensity $\left \langle |g_M(\mathbf{u_o},\mathbf{u_i},\omega)|^2 \right \rangle$ can be expressed as \cite{akkermans}:
\begin{equation}
\label{g1}
\left \langle |g_M(\mathbf{u_o},\mathbf{u_i},\omega)|^2 \right \rangle \simeq \int_{\Omega} d \mathbf{r_1} \int_{\Omega}  d \mathbf{r_2}  \left | \left \langle g_l(\mathbf{u_o},\mathbf{r_{2}},\omega) \right \rangle \right  |^2 \Gamma(\mathbf{r_{2}},\mathbf{r_{1}}) \left | \left \langle g_l(\mathbf{r_{1}},\mathbf{u_i},\omega)\right \rangle \right  |^2
\end{equation}
where $\Omega$ denotes the volume occupied by the scattering medium in the FOV of the MO.  Equation (\ref{g1}) can be given a physical interpretation by reading the integrands from right to left. The term $\left | \left \langle g_l(\mathbf{r_{1}},\mathbf{u_i},\omega) \right \rangle \right  |^2$ accounts for the intensity of the incident wave-field at point $\mathbf{r_1}$ where the first scattering event occurs. $\Gamma(\mathbf{r_{2}},\mathbf{r_{1}})$ corresponds to the impulse response for the energy transport from $\mathbf{r_{1}}$ to $\mathbf{r_{2}}$ where the last scattering event occurs [see Fig.~\ref{fig2}(b)]. $ \left | \left \langle  g(\mathbf{u_o},\mathbf{r_2},\omega) \right \rangle \right  |^2$ describes the propagation between the last scattering event and the detector located at point $\mathbf{u_o}$ in the pupil plane. Injecting Eq.~(\ref{gL}) into Eq.~(\ref{g1}) leads to
\begin{equation}
\label{g2}
\left \langle |g_M(\mathbf{u_o},\mathbf{u_i},\omega)|^2 \right \rangle = \frac{p(\mathbf{u_o}) p(\mathbf{u_i})}{(\lambda f)^4} \int_\Omega  d \mathbf{r_1} \int_\Omega d\mathbf{r_2} \exp[-(z'_1+z'_2)/\ell_s] \Gamma(\mathbf{r_2},\mathbf{r_1})~.
\end{equation}
$z'_1=(F-z_1)/\mu_i$ and $z'_2=(F-z_2)/\mu_o$ are the distances traveled by the incoming and out-coming waves within the scattering sample before and after the first and last scattering events, respectively. The integral in the above equation is directly related to a physical quantity $\alpha$ called the incoherent \textit{albedo} (or bistatic coefficient) in diffusion theory \cite{akkermans,Akkermans2}. It is defined as the ratio of the energy flux per unit solid angle $d\Omega$ to the incident energy flux. Its expression is given by:
\begin{equation}
\label{alpha}
\alpha =\frac{1}{(4 \pi)^2 W^2} \int_\Omega d\mathbf{r_1} \int_\Omega d\mathbf{r_2} \exp[-(z'_1+z'_2)/\ell_s]  \Gamma(\mathbf{r_2},\mathbf{r_1}) ~.
\end{equation}
The radiative transfer equation yields the following expression for the bistatic coefficient $\alpha $ \cite{akkermans}:
\begin{equation}
\label{static_albedo}
\alpha (\mu_r) \sim  \frac{3}{4\pi} \mu_r \left ( \mu_r + \frac{z_0}{\ell_t} \right ) \left ( 1 - \frac{5}{3 }\frac{\ell_t}{L+2z_0} \right ) 
\end{equation}
with $\theta_r=\theta_o-\theta_i$, the reflection angle and $\mu_r=\cos(\theta_r)$, its director cosine. $z_0$ is the distance from the scattering layer boundary at which the energy flux should cancel according to diffusion theory. In presence of a refractive index mismatch at the scattering medium boundaries, it can be expressed as \cite{Xhu}
\begin{equation}
z_0=\frac{2}{3}\ell_t\frac{1+R}{1-R}
\end{equation}
with $R$ the internal reflection coefficient. In the following of the study, for analytical tractability, the albedo $ \alpha(\mu) $ will be considered as constant with respect to the direction cosine $\mu$ and replace by its average $\alpha$ over the NA of the MO:
\begin{equation}
\alpha \simeq \frac{1}{2\sin^2(\beta/2)} \int_{\cos \beta }^1 d \mu_r \alpha(\mu_r) 
\end{equation}
with $\beta=\arcsin (NA)$ the acceptance angle of the MO. Under this approximation and using Eq.~(\ref{alpha}), Eq.~(\ref{g2}) can be rewritten as,
 \begin{equation}
\label{g3}
\left \langle |g_M(\mathbf{u_o},\mathbf{u_i},\omega)|^2 \right \rangle =  \frac{(4 \pi)^2 \alpha W^2}{(\lambda f)^4}   p(\mathbf{u_o}) p(\mathbf{u_i})~.
\end{equation}
This expression yields the MS intensity in the pupil area but it does not provide any information on its coherence length. According to the van-Cittert Zernike theorem \cite{goodman_stat}, the mutual coherence of the wave-field in the back focal plane should correspond to the Fourier transform of the intensity distribution in the object focal plane. By assuming this intensity distribution to be uniform over the FOV $W^2$, the coherence length $l_c$ of the impulse response in the pupil plane $g_M(\mathbf{u_o},\mathbf{u_i},\omega)$ is then given by
\begin{equation}
\label{lc}
l_c \sim \frac{\lambda f}{W}.
\end{equation}
Typically, $l_c$ is of the order of $\delta_r=\lambda/(2 NA)$, the resolution length of the imaging system.

\subsection{Single-to-multiple scattering ratio for a point-like target}

Now that both the SS and MS components of the impulse response have been expressed in the pupil plane, we can derive the corresponding SMR. The mean SS intensity is constant and equals to $|\langle g_S(\mathbf{u_o},\mathbf{u_i},\omega) \rangle|^2$ [Eq.~(\ref{eq2})]. Assuming the MS intensity is incoherent in the pupil plane, this contribution can be estimated by $\left \langle |g_M(\mathbf{u_o},\mathbf{u_i},\omega)|^2\right \rangle$ [Eq.~(\ref{g3})]. The SMR in the pupil plane, referred to as SMR$_p$, is then given by:
\begin{equation}
\label{SMR}
\mbox{$SMR_p$}=\frac{1}{(4 \pi)^2} \frac{\sigma}{W^2} \frac{e^{-\frac{F}{ \ell_s } \left (\frac{1}{\mu_i}+\frac{1}{\mu_o}\right) }}{\alpha} ~.
\end{equation}
This expression is insightful since it yields the different parameters that govern the SMR in the pupil plane. It can be expressed as the product of two ratios. The first ratio between the scattering cross-section $\sigma$ and the FOV tells us simply that the brighter or bigger the target is, the stronger the SS intensity will be. The second ratio indicates the effect of the scattering medium on the two components of the detected intensity. The SS intensity experiences an exponential attenuation while the MS background intensity is accounted for by the diffusive albedo.\\
Since both the SS and MS intensities are constant over the pupil plane, this configuration may not be best suited for an efficient separation of these two components. However, the coherence of the SS contribution can be taken advantage of to perform such a discrimination \cite{aubry2009,Kang} but it requires the measurement of the complex field by interferometric means. As we will see now, the investigation of the reflected intensity in the imaging plane offers a direct distinction between the SS and MS components. It actually corresponds to the imaging configuration of a conventional microscope.

\section{Conventional microscopy}

In this section, we want to assess theoretically the SMR for a conventional microscope. For sake of simplicity but without loss of generality, we will assume a plane wave illumination instead of the commonly adopted K\"{o}hler illumination. The latter one actually corresponds to an incoherent superposition of plane waves. We thus investigate the impulse response $g(\mathbf{r_o},\mathbf{u_i},\omega)$ between a point source $\mathbf{u_i}$ in the input pupil plane and a detector point $\mathbf{r_o}$ in the conjugate imaging plane ($z=4f$) [see Fig.~\ref{fig1}(a)]. This impulse response can be expressed as the spatial Fourier transform at the output of $g(\mathbf{u_o},\mathbf{u_i},\omega)$ \cite{goodman_fourier}, such that:
\begin{equation}
\label{eq3}
g(\mathbf{r_o},\mathbf{u_i},\omega)=\frac{e^{2ikf}}{i\lambda f} \int d^2\mathbf{u_o} g(\mathbf{u_o},\mathbf{u_i},\omega) e^{-i\frac{k}{f}\mathbf{u_o}.\mathbf{r_{o}}} ~.
\end{equation}

\subsection{Single scattering contribution}

The SS impulse response $g_S(\mathbf{r_o},\mathbf{u_i},\omega) $ is first investigated. Injecting the expression of $\langle  g_S(\mathbf{u_o},\mathbf{u_i},\omega) \rangle  $ [Eq.~(\ref{eq2})] into Eq.~(\ref{eq3}) leads to the following expression for $\langle g_S(\mathbf{r_o},\mathbf{u_i},\omega) \rangle $:
\begin{equation}
\label{eq4}
\langle g_S(\mathbf{r_o},\mathbf{u_i},\omega) \rangle = i e^{6ikf}\frac{ \sqrt{ \sigma } e^{-F/ (2 \ell_s \mu_i)}  e^{-F'/ (2 \ell_s)}   }{\delta_r^2 (\lambda f)} h \left (\mathbf{r_o} \right ).
\end{equation}
$F'$ is the effective target depth that accounts for the exponential decrease of the ballistic wave averaged over the NA, such that,
\begin{equation}
\label{effective}
e^{-F'/ (2 \ell_s)} = \frac{1}{2\sin^2(\beta/2)} \int_{\cos \beta }^1 d \mu e^{-F/ (2 \ell_s \mu)}.
\end{equation}
$h(\mathbf{r_o})$ is the normalized point-spread function of the optical system \cite{goodman_fourier}:
\begin{equation}
\label{h}
h(\mathbf{r_o})=\frac{1}{A} \int d^2\mathbf{u_o} p(\mathbf{u_o}) e^{-i\frac{k}{ f}\mathbf{u_o}.\mathbf{r_{o}}} 
\end{equation}
with 
\begin{equation}
\label{A}
A= \left(\frac{\lambda f}{ \delta_r}\right)^2 
\end{equation}
the MO pupil area. 
In conventional microscopy, the spatial distribution of the SS intensity is given by the square norm of the point spread function $h(\mathbf{r_o})$. In absence of aberrations and for a circular aperture $p(\mathbf{u_o})$, we retrieve the well-known Airy disk centered around the target location.

\subsection{Multiple scattering contribution}

Similarly, the MS component $g_M(\mathbf{r_o},\mathbf{u_i},\omega)$ can also be written as the spatial Fourier transform at the output of $g_M(\mathbf{u_o},\mathbf{u_i},\omega)$ [Eq.~(\ref{eq3})]. As shown previously, $g_M(\mathbf{u_o},\mathbf{u_i},\omega )$ is an incoherent wave-field of coherence length $l_c$ uniformly distributed over the pupil area $A$. Owing to the Fourier transform properties, $g_M(\mathbf{r_o},\mathbf{u_i},\omega )$ is also an incoherent wave-field of coherence length $\delta_r \sim \lambda f/\sqrt{A}$  [Eq.~(\ref{A})] uniformly distributed over the FOV of characteristic size $W \sim \lambda f/l_c$ [Eq.~(\ref{lc})]. Energy conservation implies an equality of the energy flux at the pupil and imaging planes such that,
\begin{equation}
W^2 \langle |g_M(\mathbf{r_o},\mathbf{u_i},\omega) |^2 \rangle  =  A \langle |g_M(\mathbf{u_o},\mathbf{u_i},\omega) |^2 \rangle 
\label{Im02}
\end{equation}
for $\mathbf{u_o}$ and $\mathbf{u_i}$ in the pupil area and $\mathbf{r_o}$ in the FOV of the imaging system. Injecting Eq.~(\ref{g3}) into Eq.~(\ref{Im02}) yields the following expression for $\langle |g_M(\mathbf{r_o},\mathbf{u_i},\omega) |^2 \rangle $:
\begin{equation}
\label{gmc}
\langle |g_M(\mathbf{r_o},\mathbf{u_i},\omega) |^2 \rangle =  \frac{(4\pi)^2 \alpha }{\delta_r^2 (\lambda f)^2} p(\mathbf{u_i}).
\end{equation}
This expression indicates that, for a plane wave illumination in the entrance pupil plane, the MS intensity presents an uniform distribution in the imaging plane.

\subsection{Single-to-multiple scattering ratio}

The SMR in conventional microscopy, referred to as SMR$_{\mbox{m}}$, can be expressed as the ratio between the target intensity given by $|\langle g_S(\mathbf{0},\mathbf{u_i},\omega) \rangle |^2$ [Eq.~(\ref{eq3})] and the MS background. By considering its uniform distribution in the imaging plane, this quantity is provided by $\left \langle |g_M(\mathbf{r_o},\mathbf{u_i},\omega)|^2\right \rangle$  [Eq.~(\ref{gmc})]. the averaged value of the SMR in conventional microscopy is then given by
\begin{equation}
\label{SMRm}
\mbox{SMR}_{m}=\frac{S}{(4\pi)^2}\frac{ \sigma}{\delta_r^2} \frac{e^{-2F'/\ell_s }  }{\alpha}
\end{equation}
with $S=|h(\mathbf{0})|^2$, the Strehl ratio ranging from 0, in the case of a fully aberrated point-spread function, to 1 in the ideal case \cite{Mahajan}. Compared to the previous result in the pupil plane [Eq.~(\ref{SMR})], the SMR in conventional microscopy is increased by a factor $N=(W/\delta_r)^2$ which corresponds to the number of resolution cells in the FOV. Whereas the MS background results from the incoherent superimposition of $N$ independent speckle grains, the target image results from the constructive interference of the ballistic photons over the NA. However, the aberration undergone by the target ballistic wave-front across the scattering layer degrades the target focal spot and lowers its intensity. This is accounted for by the Strehl ratio $S$ in Eq.~(\ref{SMRm}).
\begin{figure}[!ht]
\begin{center}
\includegraphics[width=1\textwidth]{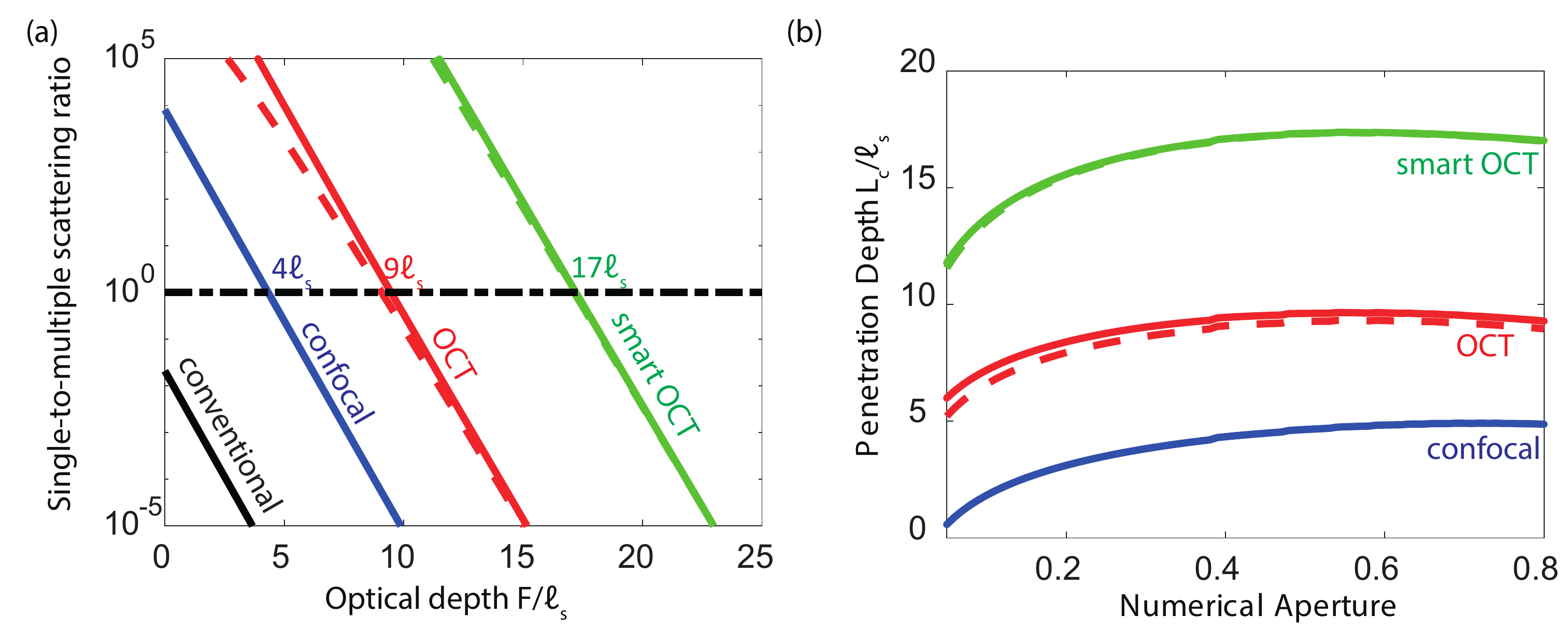}
\caption{ \label{fig3_a} (a) Single-to-multiple scattering ratio as a function of the optical depth $F/\ell_s$ for a point-like target. The performances of conventional microscopy (black line), confocal microscopy (blue line), OCT (red line) and smart OCT (green line) are compared. The y-axis is in log-scale. These curves have been computed considering experimental parameters typical of full-field OCT (see Table \ref{table}). The detection threshold (SMR$\sim1$, black dashed horizontal line) yields an imaging depth limit of $\sim 4\ell_s$ for confocal microscopy, $\sim 9 \ell_s$ for OCT and $\sim 17 \ell_s$ for smart-OCT. (b) Evolution of the corresponding penetration depths derived for each imaging technique as a function of the NA of the MO. In both (a) and (b), the SMRs in OCT and smart-OCT have been computed by considering either the diffusive approximation [Eq.~(\ref{albedo_t2})] or a Monte-Carlo calculation of the time-dependent albedo (continuous and dashed lines, respectively).}
\end{center}
\end{figure}

Figure \ref{fig3_a}(a) displays the evolution of the SMR for conventional microscopy as a function of the optical depth $F/\ell_s$. The parameters used for the computation of the SMR are described in Table \ref{table}. The considered transport parameters ($\ell_s$, $\ell_t$) are typical of in-vivo cortex tissues \cite{Schott}. The experimental parameters ($\lambda$, $F$, NA, $W$) are typical of full-field OCT \cite{Assayag}. The internal reflection coefficient $R$ is assumed to be zero since the use of an immersion microscope objective will limit the impedance mismatch with tissues. The scattering cross-section is arbitrarily chosen as the geometrical cross-section of a 1 $\mu$m diameter bead. At last, the Strehl ratio $S$ in brain tissues is estimated from a two-photon microscopy experiment that reports a fivefold signal enhancement when optical aberrations from the brain tissues are corrected with adaptive optics \cite{Ji}. Figure \ref{fig3_a}(a) shows that the depth evolution of the SMR is dominated by the exponential decrease of the ballistic wave across the scattering medium. In this example, the scattering strength of the target is too weak to obtain a SMR superior to unity. MS is predominant and prevents from a clear detection of the target. This explains why conventional microscopy can only image extremely thin layers of biological tissues (10-$\mu$m thick) and cannot allow in-vivo imaging of soft tissues.
\begin{table}[h!]
\center
 \caption{\label{table}Experimental parameters used for the theoretical prediction of the single-to-multiple scattering ratio.}
\begin{tabular}{c|c|c|c|c|c|c|c|c}
$n$ &  $\ell_s$ [$\mu$m] & $\ell_t$ [$\mu$m]& $\sigma$ [m$^2$]  & $\lambda$ [nm] & NA & S & $W$ [$\mu$m] & $\tau_c$ [fs] \\
\hline
1.4 \cite{Jacques} & 150 \cite{Schott} & 1500 \cite{Schott} &  $\pi \times 10^{-12} $   & 810  & 0.4 & 0.4 & $10^3$ & 5 \\
\end{tabular}
\end{table}

\section{Confocal microscopy}

Imaging deeper into scattering media implies a rejection of the MS background. In confocal microscopy, this rejection is performed in the spatial domain into two steps. First, an incident wave-field is focused at a given location in the focal plane. Therefore, the incident wave-field can be seen as originating from a point source $\mathbf{r_i}$ in the conjugate source plane. Secondly, the backscattered intensity is measured in the imaging plane with a detector placed after a pinhole. The corresponding impulse response $g(\mathbf{r_o},\mathbf{r_i},\omega)$ should thus be investigated. It can be expressed as a spatial Fourier transform at emission of $g(\mathbf{r_o},\mathbf{u_i},\omega)$:
\begin{equation}
\label{eq6}
g(\mathbf{r_o},\mathbf{r_i},\omega)=\frac{e^{2ikf}}{i\lambda f} \int d\mathbf{u_i} g(\mathbf{r_o},\mathbf{u_i},\omega) e^{-i\frac{2\pi}{\lambda f}\mathbf{u_i}.\mathbf{r_i}} .
\end{equation}

\subsection{Single scattering contribution}

The SS component $\langle g_S(\mathbf{r_o},\mathbf{r_i},\omega) \rangle $ is first investigated. Injecting the expression of $\langle g_S(\mathbf{r_o},\mathbf{u_i},\omega) \rangle $ [Eq.~(\ref{eq4})] into Eq.~(\ref{eq6}) leads to the following expression for $ \langle g_S(\mathbf{r_o},\mathbf{r_i},\omega)\rangle$:
\begin{equation}
\label{gbr}
\langle g_S(\mathbf{r_o},\mathbf{r_i},\omega) \rangle =  e^{8ikf}\frac{  \sqrt{ \sigma} e^{-F'/\ell_s}}{\delta_r^4 }  h (\mathbf{r_o} ) h (\mathbf{r_i} )~.
\end{equation}
In the SS regime, the impulse response between the source plane and the imaging plane corresponds to the product of the input and output point-spread functions centered around the target location, weighted by the target cross-section and the ballistic attenuation. 


\subsection{Multiple scattering contribution}

As for the SS impulse response, $g_M(\mathbf{r_o},\mathbf{r_i},\omega)$ can be written as the Fourier transform at the output of $g_M(\mathbf{r_o},\mathbf{u_i},\omega)$. Invoking the same energy conservation argument that has led to Eq.~(\ref{Im02}), the following relation can be derived:
\begin{equation}
W^2 \left \langle | g_M(\mathbf{r_o},\mathbf{r_i},\omega)|^2 \right \rangle = A \langle |g_M(\mathbf{r_o},\mathbf{u_i},\omega) |^2 \rangle ~.
\end{equation}
Injecting Eqs.~(\ref{gmc}) and (\ref{lc}) into the last equation finally yields
\begin{equation}
 \left \langle \left| g_M(\mathbf{r_o},\mathbf{r_i},\omega)\right|^2 \right \rangle= \frac{ (4\pi)^2   }{\delta_r^4 W^2} \alpha~.
\label{Imc2}
\end{equation} 
This expression indicates that the multiple scattering intensity is uniformly distributed in the imaging plane \cite{badon2016smart}.

\subsection{Single-to-multiple scattering ratio}

As previously mentioned, a confocal microscope relies on the insertion of a pinhole in the imaging plane at a position $\mathbf{r_o}$ conjugated with the incident focusing point $\mathbf{r_i}$. The confocal signal associated with the target is thus given by the single scattering component of the impulse response at the target position, namely $\langle  g_S(\mathbf{0},\mathbf{0},\omega) \rangle $. The SMR in confocal microscopy, referred to as SMR$_{\mbox{c}}$, can be expressed as the ratio between the target intensity $|\langle g_S(\mathbf{0},\mathbf{0},\omega) \rangle |^2$ [Eq.~(\ref{gbr})] and the uniform MS background given by $\left \langle |g_M(\mathbf{r_o},\mathbf{r_i},\omega)|^2\right \rangle$  [Eq.~(\ref{Imc2})], such that:
\begin{equation}
\label{SMRc}
\mbox{SMR}_{\mbox{c}}=\frac{S^2}{(4\pi)^2}\frac{ \sigma}{\delta_r^2} \left ( \frac{ W}{\delta_r} \right)^2  \frac{ e^{{ -{2F'}/{\ell_s}}}}{\alpha} ~.
\end{equation}
Compared to conventional microscopy [Eq.~(\ref{SMRm})], the SMR is increased by a factor $S \times N$. The multiplication of the SMR by $N=(W/\delta_r)^2$ results from the coherent summation of the incident ballistic wave-front at the target location. This positive effect on the SMR is reduced by the Strehl ratio $S$ which accounts for the aberration effects undergone by the incident wave-front. Figure \ref{fig3_a}(a) illustrates the benefit of the spatial rejection of MS photons performed by confocal microscopy. The corresponding SMR is displayed as a function of the optical depth $F/\ell_s$, considering the experimental parameters provided in Table \ref{table}. The SMR is drastically increased compared to conventional microscopy.

\begin{figure}[htbp]
\begin{center}
\includegraphics[width=0.9\textwidth]{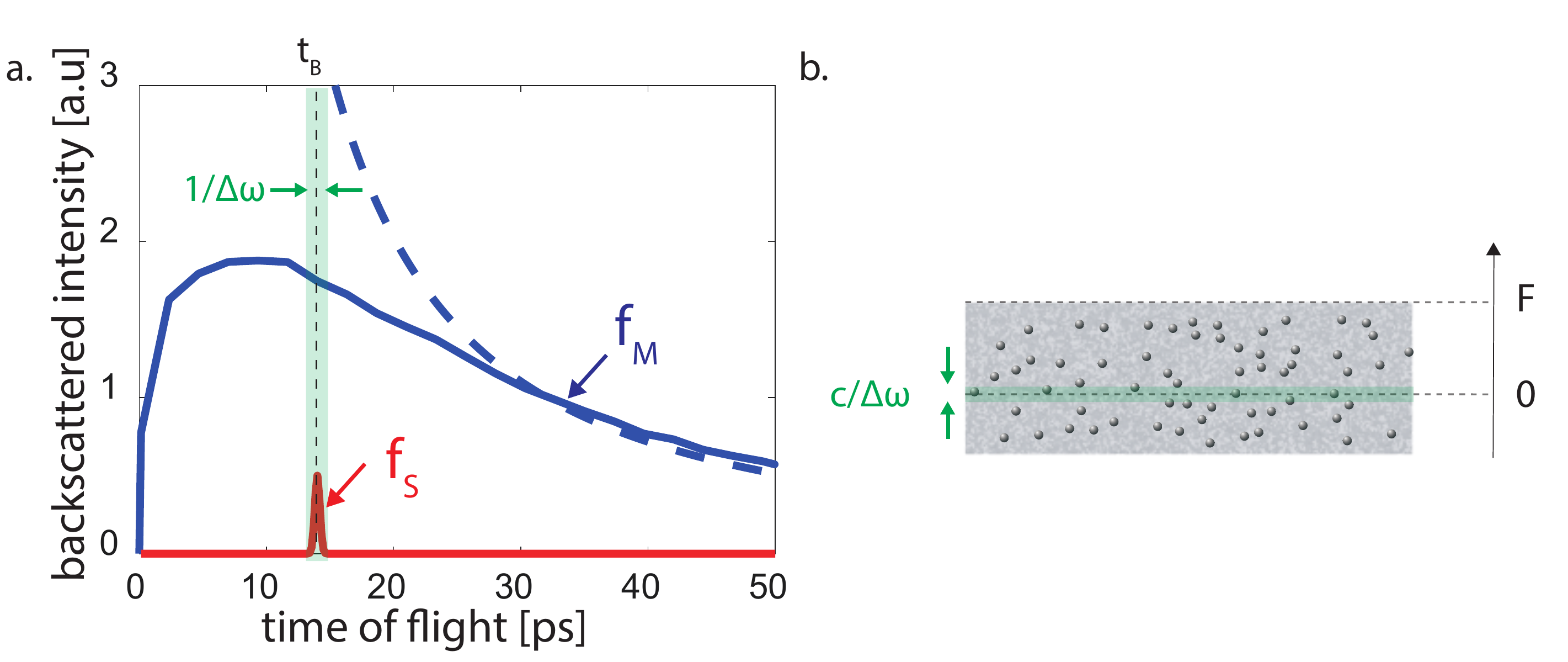}
\caption{ \label{time_dependent} (a) Time-dependent impulse response for a point-like target embedded in a scattering medium. This response can be expressed as the sum of the MS background (in blue) and the target echo arriving at time $t=t_B$ (in red). The time-of-flight distribution of MS photons can be either computed by means of a Monte-Carlo simulation (continuous line) or analytically predicted by the diffusion theory at large optical depths (dashed line, Eq.~(\ref{albedo_t2})). Coherence time gating allows to select the photons over a time window centered around this ballistic time of flight and thus drastically enhance the SMR. (b) The corresponding isochronous volume is a slab of thickness $c/\Delta \omega$ centered around the focal plane.}
\end{center}
\end{figure}

From the SMR, the penetration limit $L_c$ can be defined as the optical depth at which the SMR reaches a detection threshold of the order of unity. The corresponding imaging depth limit is displayed as a function of NA$=\lambda/(2\delta_r)$ in Fig.~\ref{fig3_a}(b). Thanks to the spatial rejection of the MS background, confocal microscopy allows to image the target at optical depths up to 5$\ell_s$ (NA$=0.7$). Note that the scaling with $N$ of the SMR should imply a continuous increase of $L_c$ with the NA. However, this effect is counteracted by a larger effective target depth $F'$ at large NA [Eq.~(\ref{effective})]: The travel length of the ballistic wave-fronts inside the scattering layer increases with the angle of incidence. This accounts for the saturation of the penetration depth at large NA. Combined with a diffraction limited resolution, this performance explains why confocal and multi-photon microscopy are the most widely used methods for \textit{in-vitro} microscopy~\cite{ntziachristos}. However, this range of penetration depths corresponds to only a few hundreds of microns into biological tissues. To go beyond this limit, one can take advantage of the distinct temporal features displayed by the SS and MS components. Indeed, while singly-scattered photons are associated with a well-defined arrival time directly linked to the target position, the MS component exhibits a broad time-of-flight distribution due to the numerous scattering events experienced by the wave-field within the scattering medium [see Fig.~\ref{time_dependent}]. This temporal discrimination between singly and multiply scattered photons is the major improvement of OCT compared to confocal microscopy.

\section{Optical coherence tomography}

In OCT, coherence time-gating allows to select the ballistic photons over a time window centered around a chosen time of flight $t$. Here, we will focus on the situation where the point source and detector are both located in the conjugate source and imaging planes, as in confocal microscopy. This configuration is actually the most favorable in terms of SMR since it both combines a confocal and a temporal discrimination of SS photons. This is the case in optical coherence microscopy when a pinhole is placed in front of the detector \cite{kempe2}, in fiber-based systems where single-mode optical fiber serves as a pinhole aperture for both illumination and collection of light from the sample \cite{izatt}, and in full-field OCT \cite{Vabre,dubois2002,thouvenin2017} where the spatial incoherence of the light source acts as a physical pinhole \cite{karamata2}.

The source emits a broadband signal $f_0(t)$ with a bandwidth $\Delta \omega$ centered around the frequency $\omega_0$. The signal measured by the detector, denoted as $f(\mathbf{r_o},\mathbf{r_i},t)$, results from the interference between a part of the incident signal taken as a reference, $s(t)$, and the back-scattered wave-field $\psi(\mathbf{r_o},\mathbf{r_i},t)$, which can be expressed mathematically as a temporal correlation:
\begin{equation}
f(\mathbf{r_o},\mathbf{r_i},t)= \psi(\mathbf{r_o},\mathbf{r_i},t)\ast s(t)~.
\end{equation}
where the symbol $\ast$ stands for a temporal correlation. The back-scattered wave-field $\psi(\mathbf{r_o},\mathbf{r_i},t)$ can be expressed itself as the convolution of the time-dependent confocal impulse response $g(\mathbf{r_o},\mathbf{r_i},t)$ with $s(t)$, such that:
\begin{equation}
\label{convol}
f(\mathbf{r_o},\mathbf{r_i},t)= g(\mathbf{r_o},\mathbf{r_i},t) \ast s(-t) \ast s(t) 
\end{equation}
which can be decomposed in the frequency domain as
\begin{equation}
\label{decompo}
f(\mathbf{r_o},\mathbf{r_i},t)=\int d\omega g(\mathbf{r_o},\mathbf{r_i},\omega) |s(\omega)|^2 e^{i \omega t}.
\end{equation}
In the same manner as above, we now decompose the recorded signal $f$ into its SS and MS components ($f=f_S+f_M$) and investigate each term separately. 

\subsection{Single scattering contribution}

Injecting Eq.~(\ref{gbr}) into Eq.~(\ref{decompo}) leads to the following expression for $\langle f_S(\mathbf{r_o},\mathbf{r_i},t) \rangle $: 
\begin{equation}
\langle f_S(\mathbf{r_o},\mathbf{r_i},t)\rangle=\int d\omega \langle g_S(\mathbf{r_o},\mathbf{r_i},\omega) \rangle |s(\omega)|^2 e^{-i \omega t}~.
\end{equation}
Assuming that $s(\omega)$ is relatively constant over the frequency bandwidth $\Delta \omega$ and using the expression of $\langle g_S(\mathbf{r_o},\mathbf{r_i},\omega) \rangle $ [Eq.~\ref{gbr}], the following expression can be found for the SS contribution at the ballistic time $t=t_B=8f/c$:
\begin{equation}
\label{eq19}
\langle  f_S(\mathbf{r_o},\mathbf{r_i},t_B) \rangle \sim  \frac{\sqrt{\sigma} }{\delta_r^4} e^{ -F'/\ell_s} \Delta \omega |s(\omega_0)|^2 h(\mathbf{r_o}) h(\mathbf{r_i})~.
\end{equation}

\subsection{Multiple scattering contribution}

As before, a probabilistic approach should be considered to investigate the MS component of the OCT signal. Using Eq.~(\ref{decompo}), the mean time-gated MS intensity $\left \langle | f_M(\mathbf{r_o},\mathbf{r_i},t) |^2 \right  \rangle  $ can be expressed as:
\begin{equation}
\left \langle  |f_M(\mathbf{r_o},\mathbf{r_i},t) |^2 \right \rangle =  \iint  d \omega_1 d\omega_2   \left \langle g_M(\mathbf{r_o},\mathbf{r_i},\omega_1)  g_M^*(\mathbf{r_o},\mathbf{r_i},\omega_2) \right \rangle |s(\omega_1)|^2 |s(\omega_2)|^2 e^{i(\omega_2-\omega_1)t}  
\end{equation}
The right hand side of the last equation displays the frequency correlation function of the monochromatic impulse response : 
\begin{equation}
 C(\mathbf{r_o},\mathbf{r_i},\omega_2-\omega_1)=\left \langle g_M(\mathbf{r_o},\mathbf{r_i},\omega_1)  g_M^*(\mathbf{r_o},\mathbf{r_i},\omega_2) \right \rangle ~.
 \end{equation}
This quantity displays a typical width $\delta \omega$ inversely proportional to the Thouless time $\tau_D$ which defines the typical time for a multiply scattered wave to go through the scattering layer \cite{thouless}. Assuming that $s(\omega)$ is relatively constant over the frequency bandwidth $\Delta \omega$ and that this latter quantity is much larger than the correlation frequency length $\delta \omega$ \cite{genack1987}, one can rewrite the previous integral as:
\begin{equation}
\left \langle  |f_M(\mathbf{r_o},\mathbf{r_i},t) |^2 \right \rangle =  \Delta \omega | s_0(\omega) |^4 \int d\omega'    C(\mathbf{r_o},\mathbf{r_i},\omega')   e^{i \omega' t} ~.
\end{equation}
with $\omega'=\omega_2-\omega_1$. Noting that the inverse Fourier transform of $C(\mathbf{r_o},\mathbf{r_i},\omega')$ corresponds to the mean intensity of the time-dependent confocal Green's function, $ \left \langle  | g_M(\mathbf{r_o},\mathbf{r_i},t ) |^2 \right \rangle$, the last equation can be rewritten as,
\begin{equation}
\label{eq20}
\left \langle  |f_M(\mathbf{r_o},\mathbf{r_i},t) |^2 \right \rangle  \sim  \Delta \omega  | s_0(\omega_0) |^4    \left \langle  | g_M(\mathbf{r_o},\mathbf{r_i},t ) |^2 \right \rangle ~.
\end{equation}
$\left \langle | g_M(\mathbf{r_o},\mathbf{r_i},t)|^2 \right \rangle$ can be deduced from its monochromatic counterpart [Eq.~(\ref{Imc2})] by substituting the static albedo $\alpha$ by its time-dependent counterpart $\alpha(t')$ \cite{akkermans}, such that:
\begin{equation}
\label{g_D}
\left \langle | g_M(\mathbf{r_o},\mathbf{r_i},t)|^2 \right \rangle =  (4\pi)^2  \frac{ \alpha(t')}{\delta_r^4 W^2} 
\end{equation}
where $t'$ denotes the lapse time of the multiply scattered wave within the scattering medium. At the ballistic time $t=t_B$, $t'=t'_B=2F/c$. The time-dependent albedo $\alpha(t')$ is defined as the ratio of the emergent energy per unit surface, unit time and unit solid angle $d\Omega$ to the incident energy flux along a given direction. Its expression is closely related to the static albedo [Eq.~(\ref{alpha})], except that the time-dependent Green's function for energy transport, $\Gamma(\mathbf{r_2},\mathbf{r_1},t')$, is now considered  \cite{akkermans}:
\begin{equation}
\label{alphat}
\alpha(t')=\frac{1}{(4 \pi)^2 W^2} \int d\mathbf{r_1} d\mathbf{r_2} e^{-(z'_1+z'_2)/\ell_s}  \Gamma(\mathbf{r_2},\mathbf{r_1},t') ~.
\end{equation}
As the target is embedded within the scattering medium, the time-dependent albedo for a semi-infinite medium should be considered. Under the diffusive approximation, its expression is the following \cite{akkermans,Akkermans2}:
\begin{equation}
\label{albedo_t2}
\alpha(t')\simeq \frac{c(z_0+\ell_s)^2}{(4\pi Dt')^{3/2}}~.
\end{equation}
This analytical result is compared in Fig.~\ref{time_dependent}(a) with the exact calculation of the time-dependent albedo performed with a Monte-Carlo simulation. The experimental conditions are given in Table \ref{table}. There is only a quantitative agreement between the two curves at times longer than four transport mean free times ($\ell_t/c \sim 7$ ps). Nevertheless, as we will show further, the diffusive approximation yields a correct prediction of the SMR and penetration depth for OCT. As illustrated by Fig.~\ref{time_dependent}(a), coherence time gating allows to drastically reject multiply-scattered photons. Injecting Eq.~(\ref{eq20}) into Eq.~(\ref{g_D}) leads to the final expression for the time-gated MS intensity:
\begin{equation}
\label{IDoct}
\left \langle  |f_M(\mathbf{r_o},\mathbf{r_i},t) |^2 \right \rangle  \sim   \frac{(4 \pi)^2 \alpha(t')}{\delta_r^4 W^2}  \Delta \omega  | s_0(\omega_0) |^4 ~.
\end{equation}

\subsection{Single-to-multiple scattering ratio}

The SMR in OCT, referred to as SMR$_{\mbox{t}}$, can be expressed as the ratio between the target intensity $|\langle f_S(\mathbf{0},\mathbf{0},t_B) \rangle|^2$ [Eq.~(\ref{eq19})] and the MS intensity $\left \langle  |f_M(\mathbf{r_o},\mathbf{r_i},t'_B) |^2 \right \rangle $  [Eq.~(\ref{IDoct})] at the ballistic time $t_B$, such that:
\begin{equation}
\label{SMRt}
\mbox{SMR}_{\mbox{t}}=\frac{S^2}{(4\pi)^2} \frac{ \sigma }{ \delta_r^2 } \left ( \frac{W}{\delta_r} \right)^2  \frac{e^{{ -{2F'}/{\ell_s}}}\Delta \omega}{\alpha(t'_B)}~.
\end{equation}
Compared to confocal microscopy [Eq.~(\ref{SMRc})], the SMR in OCT is increased by a factor $\Delta \omega (\alpha/\alpha(t'_B))$ which accounts for the ratio of multiply scattered photons rejected by coherence time gating. Note that, in absence of any confocal filter in the OCT system, the SMR can also be derived by multiplying, this time, the SMR previously derived for conventional microscopy [Eq.~(\ref{SMRm})] with the factor $\Delta \omega (\alpha/\alpha(t'_B))$. 

The broader the source bandwidth is, the more efficient the discrimination is. Coherence time gating is usually performed with a pulse laser \cite{fujimoto} or with thermal light source in the case of full-field OCT \cite{Vabre,dubois2002,thouvenin2017}, resulting in a coherence time $\tau_c=\Delta \omega^{-1}$ of a few femtoseconds in the latter case. Figure \ref{fig3_a}(a) illustrates the benefit of a temporal discrimination of singly-scattered photons by showing the SMR evolution with the optical depth $F/\ell_s$ for OCT. The considered experimental parameters, in particular the coherence time $\tau_c$, are provided in Table \ref{table}. The SMR is drastically increased compared to conventional and confocal microscopy. From the SMR, an imaging depth limit can be deduced as the optical depth at which the SMR reaches a detection threshold of the order of unity. The corresponding penetration depth limit is displayed as a function of the NA in Fig.~\ref{fig3_a}(b). Thanks to the temporal discrimination of the SS component, OCT allows to image the target at optical depths from 6$\ell_s$ (NA$=0.05$) to 9.5$\ell_s$ (NA$=0.5$). The comparison with confocal microscopy illustrates the benefit of coherence time gating. This confirms that OCT is currently the most mature optical microscopic technique.

Interestingly, in a recent paper \cite{badon2016smart}, a matrix approach of OCT has been proposed to push back this fundamental limit of MS. Experimentally, this approach, referred to as smart-OCT, relies on the measurement of a time-gated reflection matrix from the scattering medium. An input-output analysis of the reflection matrix allows to get rid of most of the MS contribution. Iterative time-reversal \cite{prada,popoff3} is then applied to overcome the residual MS contribution as well as the aberration effects induced by the turbid medium itself. In terms of SMR, the smart-OCT approach allows to improve the OCT performance by a factor $N$, the number of spatial degrees of freedom provided by the imaging system. This leads to an extension of the imaging-depth limit by almost a factor two compared to standard OCT [see Fig.~\ref{fig3_a}].

\section{Imaging of an extended target}

In the previous section we have derived the expression of the SMR for various imaging methods in the case of an individual scatterer. However, extended reflectors, such as a United States Air Force (USAF) target, are often used as figure of merit to demonstrate performances of imaging techniques in scattering media. We now propose to investigate this configuration which, as we will see, can notably change the expected performances of the different imaging methods.
\begin{figure}[htbp]
\begin{center}
\includegraphics[width=\textwidth]{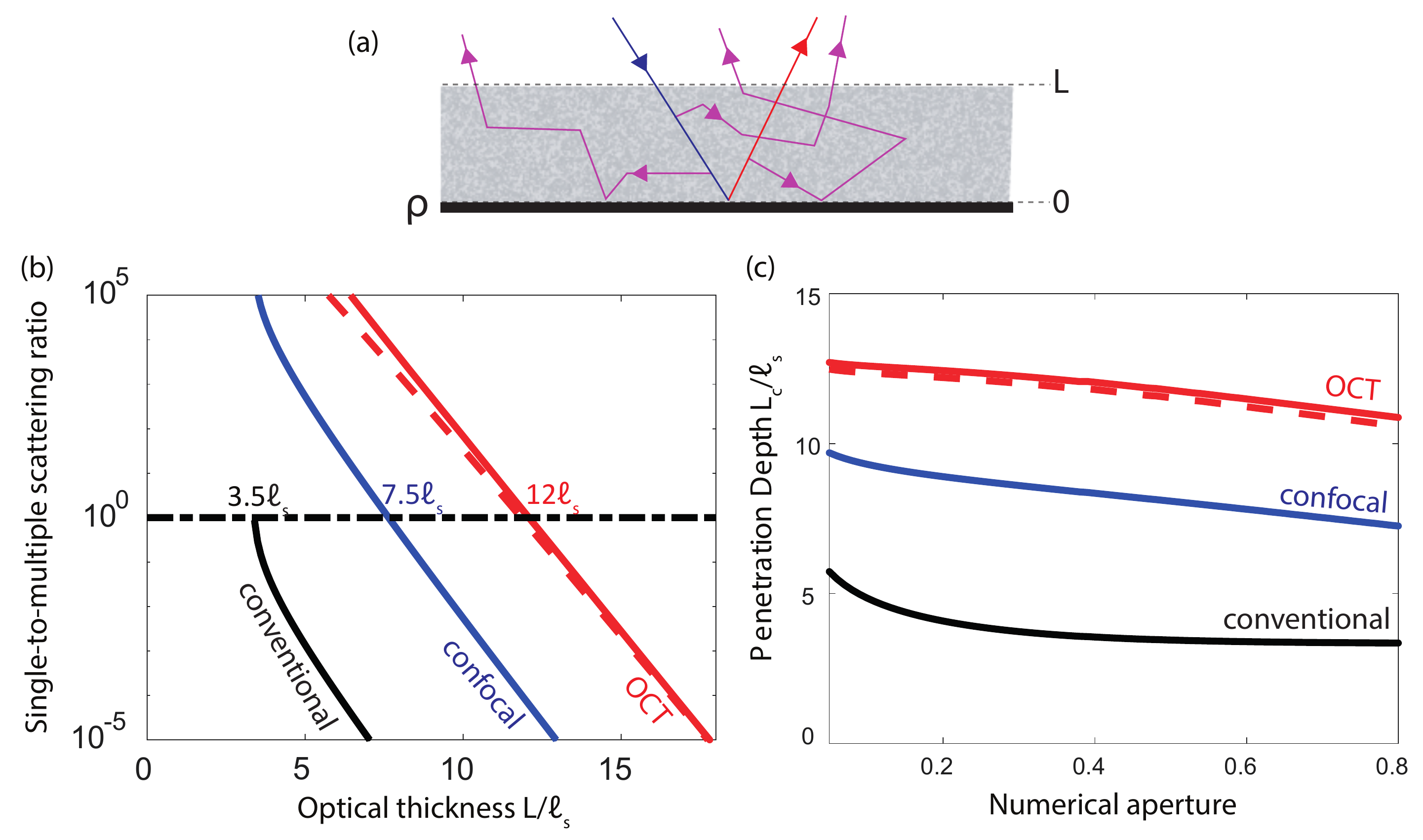}
\caption{ \label{fig_extended} (a) Scheme of an extended target placed behind a scattering layer of thickness $L$. (b) Single-to-multiple scattering ratio as a function of the optical thickness $L/\ell_s$ for an extended target. The performances of conventional microscopy (black line), confocal microscopy (blue line) and OCT (red line) are compared. The y-axis is in log-scale. These curves have been computed considering experimental parameters typical of full-field OCT (see Table \ref{table}) and for a target reflectivity $|\rho|=1$. The detection threshold (SMR$\sim1$, black dashed horizontal line) yields an imaging depth limit of $\sim 3.5\ell_s$ for conventional microscopy, $\sim 7.5\ell_s$ for confocal microscopy, $\sim 12 \ell_s$ for OCT. (c) Evolution of the corresponding penetration depths derived for each imaging technique as a function of the NA of the MO. In panels (a) and (b), the SMR in OCT has been computed by considering either the diffusive approximation [Eq.~(\ref{albedo_t2})] or a Monte-Carlo calculation of the time-dependent albedo (continuous and dashed lines, respectively).}
\end{center}
\end{figure}

\subsection{Single scattering component in the pupil plane}

For an extended target such as a United States Air Force (USAF) target or an interface between tissue layers in medical imaging, the target can be considered as a mirror of reflectivity $\rho$ [Fig.~\ref{fig_extended}(a)]. For the sake of simplicity, the target is assumed to be in the focal plane ($z=0$) and its normal $\mathbf{n}$ aligned along the optical axis. The Rayleigh-Sommerfeld integral can be used to express the reflected wave-field in the pupil plane, $g_S(\mathbf{u_o},\mathbf{u_i},\omega)$, as a function of the incident wave-field in the focal plane, $g_l(\mathbf{r_{\|}},\mathbf{u_i},\omega)$, such that:
\begin{equation}
\label{gS2}
g_S(\mathbf{u_o},\mathbf{u_i},\omega)= 2 \rho \int d^2 \mathbf{r_{\|}}  g_{l}(\mathbf{u_o},\mathbf{r_{\|}},\omega)  \partial_n g_{l}(\mathbf{r_{\|}},\mathbf{u_i},\omega) ~.
\end {equation}  
Using the expression of the lens impulse response $\langle g_l \rangle $ given in Eq.~(\ref{gL}), the ensemble average of $g_S(\mathbf{u_o},\mathbf{u_i},\omega)$ can be simplified into:
 \begin{equation}
\label{gs3}
\langle g_S(\mathbf{u_o},\mathbf{u_i},\omega) \rangle = {2 i  e^{4ikf} k \rho} e^{-L/( \ell_s \mu_o ) }  \delta_r \left (\mathbf{u_i}+\mathbf{u_o}\right )
\end {equation}
Unlike a point-like target that gives rise to a constant reflected intensity in the pupil plane [Eq.~(\ref{eq2})], a mirror-like interface implies a specular reflection: $g_S(\mathbf{u_o},\mathbf{u_i},\omega) \neq 0$ only when $\mathbf{u_i}=-\mathbf{u_o}$. 

\subsection{Conventional microscopy}

A conventional microscopy configuration is now considered with a detector point $\mathbf{r_o}$ in the imaging plane. The corresponding impulse response $g_S(\mathbf{r_o},\mathbf{u_i},\omega)$ can be expressed as a Fourier transform at reception of $g_S(\mathbf{u_o},\mathbf{u_i},\omega)$ [see Eq.~(\ref{eq3})]. It yields:
\begin{equation}
\label{gs4}
\langle g_S(\mathbf{r_o},\mathbf{u_i},\omega) \rangle = e^{6ikf} \frac{2   k \rho}{\lambda f} e^{-2 L'/\ell_s} p \left (-\mathbf{u_i} \right ) e^{i \frac{k}{f} \mathbf{u_i}.\mathbf{r_o}} 
\end{equation}
with $L'$, the effective scattering thickness, such that
\begin{equation}
\label{effective2}
e^{-L'/  \ell_s} = \frac{1}{2\sin^2(\beta/2)} \int_{\cos \beta }^1 d \mu_o e^{-L/ (\ell_s \mu_o)}.
\end{equation}
Not surprisingly, the reflected wave-field corresponds to a plane wave of parallel momentum $-{k}\mathbf{u_i}/f$ weighted by the pupil function $p \left(-\mathbf{u_i} \right )$. The corresponding SMR can be expressed as the ratio between the target intensity $|\langle g_S(\mathbf{0},\mathbf{u_i},\omega) \rangle |^2$ and the MS background $\left \langle |g_M(\mathbf{r_o},\mathbf{u_i},\omega)|^2\right \rangle$  previously derived in Eq.~(\ref{gmc}): 
\begin{equation}
\label{SMRm2}
\mbox{SMR}_{\mbox{m}}=  \left (\frac{ |\rho| }{ 2 NA } \right)^2 \frac{e^{{ -{2L'}/{\ell_s}}} }{  \alpha} ~.
\end{equation}
Contrary to its expression for a point-like target [Eq.~(\ref{SMRm})], the SMR in conventional microscopy for an extended target is not sensitive on aberrations and tends to decrease with the NA of the MO [see Fig.~\ref{fig_extended}(c)]. Figure \ref{fig_extended}(b) shows the evolution of the corresponding SMR with the optical thickness in the experimental conditions provided in Table \ref{table}. Not surprisingly, the SMR level is much higher than in the case of a point-like target [Fig.~\ref{fig3_a}(a)]. The imaging depth limit is of 3.5$\ell_s$ for an extended target. This result demonstrates why an experimental demonstration based on the imaging of a resolution target cannot be used to estimate the penetration depth of an imaging technique.
 
\subsection{Confocal microscopy}

The confocal microscopy configuration can also be addressed by means of a Fourier transform of the wave-field this time at the input [see Eq.~(\ref{eq6})]. The corresponding impulse response $\langle  g_S(\mathbf{r_o},\mathbf{r_i},\omega) \rangle $ is then given by
\begin{equation}
\label{gs5}
\langle g_S(\mathbf{r_o},\mathbf{r_i},\omega) \rangle = e^{8ikf} \frac{2 k \rho e^{-L'/\ell_s}}{\delta_r^2}  h \left (\mathbf{r_o}-\mathbf{r_i} \right) . 
\end{equation}
Interestingly, the SS component does not depend on the product of the input and output point-spread function [Eq.~(\ref{gbr})] but on a single point-spread function $h \left (\mathbf{r_o}-\mathbf{r_i} \right)$. The corresponding SMR can be expressed as the ratio between the target intensity $|\langle  g_S(\mathbf{0},\mathbf{0},\omega)\rangle |^2$ and the MS background $\left \langle |g_M(\mathbf{r_o},\mathbf{r_i},\omega)|^2\right \rangle$  previously derived in Eq.~(\ref{Imc2}):
\begin{equation}
\label{SMRc2}
\mbox{SMR}_{\mbox{c}}= S |\rho|^2 \left (\frac{ W }{ \lambda } \right)^2 \frac{e^{{ -{2L'}/{\ell_s}}} }{  \alpha} ~.
\end{equation}
Contrary to its expression for a point-like target [Eq.~(\ref{SMRc})], the SMR for an extended target scales as $S$ (and not as $S^2$). Not surprisingly, confocal microscopy is thus less sensitive to aberrations for an extended target. More interestingly, the extended size of target implies a SMR scaling as $(W/\lambda)^2$ instead of $(W/\delta_r)^2$ [see Eqs.~(\ref{SMRc}) and (\ref{SMRc2})]. Unlike the point-like target case, the confocal gain does not depend on the NA for an extended target. Physically, this can be understood as follows: An extended target will back-reflect all the incident ballistic energy whatever the NA, whereas a point-like target will only scatter a fraction of it ($\sim \sigma/\delta_r^2 $, see Eq.~(\ref{SMRc})). For an extended target, the imaging performance thus slightly decreases with the NA [see Fig.~\ref{fig_extended}(c)] because the effective optical thickness $L'$ increases at larges angles of incidence and reflection. Figure \ref{fig_extended}(b) shows the evolution of the corresponding SMR with the optical thickness in the experimental conditions provided in Table \ref{table}. The optical thickness limit for confocal microscopy is here of about 7.5$\ell_s$, which is almost two times deeper than for a point-like target [see Fig.~\ref{fig3_a}(a)]. Again, this difference shows why an experimental demonstration based on the imaging of a resolution target cannot be used to estimate the penetration depth of a confocal microscope.

\subsection{Optical coherence tomography}

With regards to OCT, the time-gated target signal $\langle f_S(\mathbf{r_o},\mathbf{r_i},t_B) \rangle $ can be derived by injecting the expression of $\langle g_S(\mathbf{r_o},\mathbf{r_i},\omega) \rangle $ [Eq.~(\ref{gs5})] into Eq.~(\ref{decompo}). It yields
\begin{equation}
\label{eq19b}
\langle f_S(\mathbf{r_o},\mathbf{r_i},t_B) \rangle \sim  \frac{2 \rho k}{\delta_r^2} e^{{ -{L'}/{\ell_s}}}  \Delta \omega |s_0(\omega_0)|^2 h(\mathbf{r_o}-\mathbf{r_i})
\end{equation}
The SMR in OCT is then deduced by considering the ratio between the target intensity $|\langle f_S(\mathbf{0},\mathbf{0},t_B) \rangle|^2$ [Eq.~(\ref{eq19b})] and the MS intensity $\left \langle  |f_M(\mathbf{r_o},\mathbf{r_i},t'_B) |^2 \right \rangle $  [Eq.~(\ref{IDoct})] at the ballistic time $t_B$, such that
\begin{equation}
\label{SMRt2}
\mbox{SMR}_{\mbox{t}}=S |\rho|^2  \left (\frac{ W }{ \lambda } \right)^2 \frac{   e^{{ -{2L'}/{\ell_s}}} \Delta \omega }{ \alpha(t'_B)} 
\end{equation}
As for confocal microscopy,  the extended size of target implies a SMR scaling as $(W/\lambda)^2$ and $S$ instead of $(W/\delta_r)^2$ and $S^2$ for a point-like target [see Eqs.~(\ref{SMRt2}) and (\ref{SMRt})]. Therefore, the penetration limit slightly decreases with the NA of the MO for an extended target [see Fig.~\ref{fig_extended}(c)], in contrast with the point-like target case [see Fig.~\ref{fig3_a}(b)]. Figure \ref{fig_extended}(b) shows the evolution of the corresponding SMR with the optical thickness in the experimental conditions provided in Table \ref{table}. Coherence time gating allows to push back the optical thickness limit to 12$\ell_s$, instead of 7.5$\ell_s$ for confocal microscopy. This theoretical value is in agreement with the imaging-depth limit recently reached by Kang \textit{et al.} \cite{Kang} with a resolution target.

\section{Imaging of soft tissues}
Most optical microscopy techniques aim at imaging soft tissues themselves rather than just detecting a bright target through them. In this last section, we thus address this important issue for medical applications. 

\begin{figure}[htbp]
\begin{center}
\includegraphics[width=0.5\textwidth]{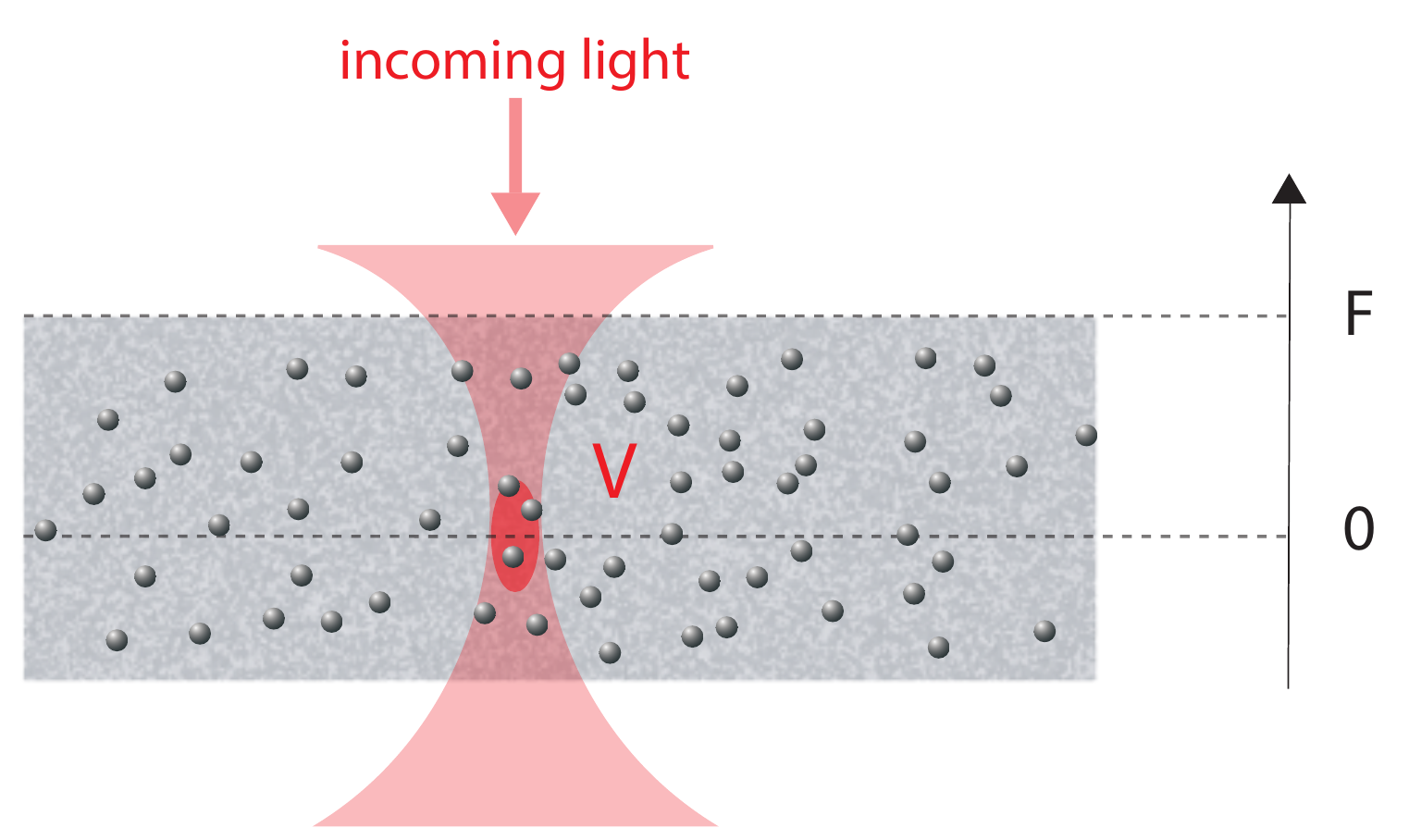}
\caption{ \label{figure_tissues} Modeling of soft tissues. The biological media are made of randomly distributed scatterers at a concentration $n$. The brightness of tissues can be quantified by the back-scattering cross-section $\sigma_b $[Eq.~(\ref{sigma_random2})]. }
\end{center}
\end{figure}
In biological media, the SS signal results from the incoherent summation of the echoes due to scatterers randomly distributed inside the volume $V$ of the focal spot  [see Fig.~\ref{figure_tissues}]. In an isotropic medium, the corresponding overall scattering cross-section $\sigma$ is given by
\begin{equation}
\label{sigma}
\sigma= 4 \pi n \frac{d \sigma_s}{d \Omega} V=n \sigma_s V  
\end{equation}
$n$ is the concentration of scatterers, $\frac{d \sigma_s}{d \Omega}$ is the differential scattering cross-section of each scatterer and $\sigma_s$ their scattering cross-section. In the dilute regime, the scattering mean free path $\ell_s$ is directly equal to $(n\sigma_s)^{-1}$ \cite{akkermans}, which yields the following relation
\begin{equation}
\label{dilute}
\sigma=\frac{ V}{\ell_s}
\end{equation}

However, the scattering of soft tissues is generally highly anisotropic. To take into account the weakness of light back-reflection in biological media, a back-scattering cross-section $\sigma_b$ is thus introduced to replace $\sigma$ [Eq.~(\ref{sigma})]:
\begin{equation}
\label{sigma_random}
\sigma_b = 4 \pi n \left \langle \frac{d \sigma}{d \Omega} \right \rangle_{NA} V  
\end{equation}
where $ \langle d\sigma/d \Omega  \rangle_{NA}  $ is  the average of the differential scattering cross-section $d\sigma/d \Omega $  over the NA of the MO:
\begin{equation}
\label{diff}
 \left \langle \frac{d \sigma}{d \Omega} \right \rangle_{NA}  = \frac{1}{2\sin^2(\beta/2)} \int_{0}^{\beta} d\theta \frac{d \sigma(\theta) }{d \Omega}\sin \theta
\end{equation}
In analogy with the isotropic case [Eq.~(\ref{dilute})], one can define a \textit{back-scattering} mean free path, such that:
\begin{equation}
\label{sigma_random2}
\sigma_b = \frac{V}{\ell_b} ~.
\end{equation}
In soft tissues, the back-scattering mean free path $\ell_b$ is of the order of transport mean free path $\ell_t$.

\subsubsection{Conventional and confocal microscopy}

In conventional or confocal microscopy, the volume $V$ of the focal spot is $p \delta_r^2 $ with $p$ the depth of field of the MO \cite{born}:
\begin{equation}
\label{p}
p=2 n\lambda/( NA)^2
\end{equation}
The backscattering cross-section $\sigma_b$ of the focal spot [Eq.~(\ref{sigma_random2})] is thus given by: 
\begin{equation}
\label{sigma_random3}
\sigma_b = \frac{p \delta_r^2}{\ell_b}~.
\end{equation}
By replacing $\sigma$ by $\sigma_b$ into Eq.~(\ref{SMRm}), the SMR in conventional microscopy can be expressed as:
\begin{equation}
\label{SMRm3}
\mbox{SMR}_{\mbox{m}}=\frac{S}{(4\pi)^2}\frac{p}{\ell_b} \frac{ e^{{ -{2F'}/{\ell_s}}}}{\alpha}~.
\end{equation}
Compared to the case of a point-like target [Eq.~(\ref{SMRm})], the SMR in conventional microscopy increases with the depth-of-field $p$ and thus decreases with the NA in soft tissues. Decreasing the NA seems more favorable to mitigate the effect of MS but this is of course at the cost of a loss in resolution. Anyhow, the penetration depth is extremely shallow in biological tissues for conventional microscopy.

\begin{figure}[!ht]
\begin{center}
\includegraphics[width=1\textwidth]{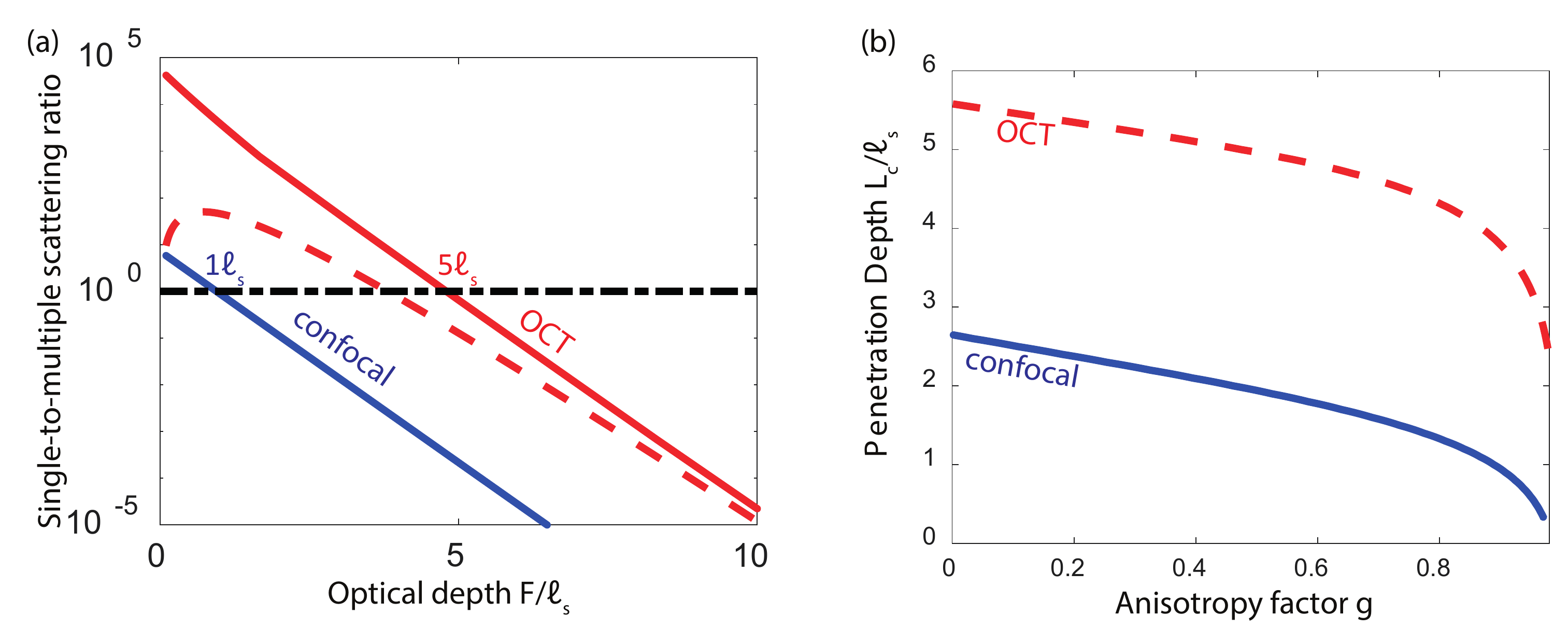}
\caption{ \label{fig3} Imaging depth limit in biological tissues considering experimental parameters typical of full-field OCT (see Table \ref{table}). (a) Single-to-multiple scattering ratio as a function of the optical thickness $F/\ell_s$ . The performances of confocal microscopy (blue line) and OCT (red line) are compared. The SMR in OCT has been computed by considering either the diffusive approximation [Eq.~(\ref{albedo_t2})] or a Monte-Carlo calculation of the time-dependent albedo (continuous and dashed lines, respectively). (b) Evolution of the penetration depths derived for confocal microscopy (blue continuous line) and OCT (diffusive prediction, dashed line) as a function of the anisotropy factor $\eta$.  } 
\end{center}
\end{figure}

The confocal SMR can also be derived by injecting the expression of $\sigma_b$ [Eq.~(\ref{sigma_random3})] into Eq.~(\ref{SMRc}),
\begin{equation}
\label{SMRc3}
\mbox{SMR}_{\mbox{c}}=\frac{S^2}{(4\pi)^2}\frac{ p}{\ell_b} \left ( \frac{ W}{\delta_r} \right)^2  \frac{ e^{{ -{2F'}/{\ell_s}}}}{\alpha} = \frac{ n^3 S^2}{2\pi^2}  \frac{ W^2}{\lambda \ell_b}   \frac{ e^{{ -{2F'}/{\ell_s}}}}{\alpha}  ~. 
\end{equation}
Compared to the confocal imaging of point-like or targets [Eqs.~(\ref{SMRc}) and (\ref{SMRc2})], the confocal SMR in soft tissues should not depend on the NA. The confocal elimination of the MS background, that scales as $N=(W/\delta_r)^2$, is counterbalanced with a shallower depth-of-field $p$ [Eq.~(\ref{p})] that implies a smaller SS contribution. Figure \ref{fig3}(a) displays the corresponding SMR as a function of the optical depth $F/\ell_s$ in the experimental conditions provided in Table \ref{table}. The differential scattering cross-section [Eq.~(\ref{diff})] is described by a Henyey-Greenstein phase function. Compared to the point-like target case, the penetration depth of confocal microscopy is here much smaller ($1\ell_s$) because of the weak back-reflection in soft tissues. Figure \ref{fig3}(b) confirms that observation by displaying the evolution of the penetration depth with the anisotropy factor $\eta=1-\ell_s/\ell_t$ for $NA=0.4$. Whereas the penetration depth is of almost $3\ell_s$ in the isotropic case ($\eta=0$), it goes below $1\ell_s$ for an extreme anisotropic case ($\eta=0.99$).

\subsubsection{Optical coherence tomography}

In OCT, the axial resolution is no longer determined by the depth of field $p$ of the MO but by the coherence time $\tau_c=(\Delta \omega)^{-1}$ of the light source such that $V= c \tau_c \delta_r^2 /2 $. The scattering cross-section of scatterers contained in the same voxel is thus given by:
\begin{equation}
\label{sigma_random4}
\sigma = \frac{c \delta_r^2}{2\ell_b \Delta \omega}~.
\end{equation}
By injecting this last expression of $\sigma$ into Eq.~(\ref{SMRt}), the SMR in OCT can be expressed as:
\begin{equation}
\label{SMRt3}
\mbox{SMR}_{\mbox{t}}=\frac{S^2}{(4\pi)^2} \frac{ c }{ 2 \ell_b } \left ( \frac{W}{\delta_r} \right)^2 \frac{e^{{ -{2F'}/{\ell_s}}} }{\alpha(t'_B)}~.
\end{equation}
Contrary to previous expressions of the SMR in OCT for a point-like or an extended target [Eqs. (\ref{SMRt}) and (\ref{SMRt2})], the SMR does not depend on the bandwidth of the light source, or equivalently on its coherence time, in a first approximation. A broader bandwidth for the light source improves the axial resolution of the image but does not increase the SMR when imaging a random distribution of scatterers. This remains true as long as the time-dependent albedo can be considered as constant over the time gate. As already pointed out experimentally \cite{Boas}, the imaging performance does not depend on the NA. There is indeed a balance between the confocal gain $N= ( W/\delta)^2$ in Eq.~(\ref{SMRt3}) and the increase of the effective optical depth $F'$ at large NA. Figure \ref{fig3}(a) displays the corresponding SMR as a function of the optical depth $F/\ell_s$ in the experimental conditions provided in Table \ref{table}. The penetration depth for OCT reaches a value of 5$\ell_s$ for soft tissues. This imaging depth corresponds to approximately 1 mm in the case of biological tissues. This prediction is in agreement with the values found in the literature \cite{dunsby2003techniques}. Note that, for low NA, a part of the multiple scattering contribution (snake photons \cite{dunsby2003techniques}) can contribute to the OCT signal and therefore increase the penetration depth. This effect is beyond the scope of this paper but has already been studied in the literature \cite{oct}.

\section{Conclusion}

A simple analytical model is developed to describe the principle and the performances of conventional microscopy, confocal microscopy, optical coherence tomography and related methods. Our approach provides different analytical expressions of the single-to-multiple scattering ratio according to the nature of the object. Qualitatively, it provides the scaling of this key quantity with the different parameters of the imaging apparatus and the scattering properties of the turbid medium. Quantitatively, it predicts the penetration depths provided by these different imaging techniques. Our predictions are in good agreement with the experimental values found in the literature. This work will provide a useful theoretical framework for the current and forthcoming studies \cite{Kang,badon2016smart,Park2015} aiming at pushing back the multiple scattering limit in optical microscopy. 

\section*{Funding}
The authors are grateful for funding provided by LABEX WIFI (Laboratory of Excellence within the French Program Investments for the Future, ANR-10-LABX-24 and ANR-10-IDEX-0001-02 PSL*) and European Research Council (ERC Synergy HELMHOLTZ). A. B. acknowledges financial support from the French ``Direction G\'{e}n\'{e}rale de l'Armement''(DGA).

\end{document}